\documentclass[preprint,showpacs,preprintnumbers,amsmath,amssymb,prb]{revtex4}
\usepackage{graphicx}
\usepackage{dcolumn}
\usepackage{bm}
\usepackage{epsfig}
\usepackage{subfigure}
\begin{document}

\title{Anomalies in a waterlike model confined between plates}

\author{Leandro B. Krott}
\address{Instituto de F\'{i}sica, Universidade Federal do Rio 
Grande do Sul, 91501-970, Porto Alegre, Rio Grande do Sul}

\author{Marcia C. Barbosa}
\address{Instituto de F\'{i}sica, Universidade Federal do Rio 
Grande do Sul, 91501-970, Porto Alegre, Rio Grande do Sul}

\date{\today}
\begin{abstract}

Using molecular dynamic simulations we study a waterlike model confined 
between two fixed hydrophobic plates.  The  system is tested 
for density, diffusion and structural anomalous behavior and 
compared with the bulk results.  Within the 
range of confining distances we had explored we
observe that in the pressure-temperature phase
diagram the  temperature of 
maximum density (TMD line), the 
temperature of maximum and minimum diffusion occur at
lower temperatures when compared with the
bulk values. For distances between the two
layers below a certain threshold, $d\le d_c$, only two layers
of particles are formed, for  $d\ge d_c$ three or more
layers are formed. In the case of three layers the central
layer stays liquid while the contact layers crystallize. This
result is in agreement with simulations for atomistic models.
\end{abstract}
\maketitle

\section{\label{sec1}Introduction}

Water has several peculiar thermodynamic and 
dynamic properties not observed in 
other liquids. This is the case of 
the density at room pressure that has 
a maximum at $4^oC$~~\cite{Wa64,Ke75,An76} while in
most materials the density increases monotonically with 
the decrease of the temperature. In addition, between $0.1$  MPa and 
$190$ MPa water also 
exhibits an anomalous increase of 
compressibility~\cite{Sp76,Ka79} and, at atmospheric pressure, an 
increase of isobaric
heat capacity upon cooling~\cite{An82,To99}. 
Besides the thermodynamic anomalies water
also exhibits an unusual behavior
in its mobility. The diffusion coefficient that for
normal liquids increases with the 
decrease of pressure, for water it
has a maximum at $4^oC$ for $1.5\;atm$~\cite{An76,Pr87}. 
The presence of the large increase in the response function 
induced the idea of the existence of two liquid phases
and a critical point~\cite{Po92}. This
critical point is located at the 
supercooled region beyond  the line of 
homogeneous nucleation and thus cannot be experimentally measured. 
In order to circumvent this inconvenience, experiments in 
confined water were performed~\cite{Li05}. They showed that the
large increase of the specific heat it is actually a
peak that can be associated with the Widom line, the
continuation of the coexistence line beyond a critical point.

The drawback of experiments in nanoscale
confinement is that the results obtained 
do not necessarily lead to conclusions 
at the bulk level. 
Notwithstanding this disadvantage the study of 
confined water by itself is interesting since 
water is present 
in ionic channels, proteins, vesicles and 
other cellular structures under 
nanoscale  confinement. 
In order to understand the behavior of water under 
these limitations a number of experiments and simulations of 
confined water have been performed. 

Several types of confinement have been explored: experiments in cylindrical 
porous~\cite{erko_2011,Mo97,MoK99} and simulations in carbon 
nanotubes~\cite{koga_zeng_2001,hummer_2001}, simulations in porous 
matrices~\cite{gallo_2007, strekalova_2012, bonnaud_2010,pizio_domingues_2010}, 
experiments~\cite{BeF96,ZaB05} and simulations in rough surfaces~\cite{tanaka_koga_2005, Ku05, lupkowski_1990,scheidler_binder_2002, Choudhury_2010}
and simulations in flat plates~\cite{meyer_stanley_1999, Gi09, scheidler_binder_2002,Ku05}.

In particular, x-ray and neutron scattering  with 
water in nanopores
 show that the liquid 
state persists down to temperature much lower 
than in bulk~\cite{erko_2011,De10,Ja08}. In these experiments in the
case  of hydrophobic
walls the liquid-crystal transition occurs 
at lower temperatures than in the case
of hydrophilic walls~\cite{De10,faraone_liu_chen_2009}. 
In some scattering experiments there are indications of the
formation of cubic ice instead of
the hexagonal ice present in the bulk~\cite{Be93,Mo97}. Several of these experiments 
show evidences of the presence of layers, one close to the walls
and one at the center~\cite{erko_2011,De10,Ja08,Mo97}. For certain
type of walls, the central crystallizes before the 
wall layers~\cite{erko_2011,Mo97}. Therefore, the experimental
results are not conclusive. They indicate that the 
crystallization in confined 
water depends strongly on the size of the 
porous~\cite{MoK99,HwC07,Ja08,KiS09,erko_2011} 
and on the level of hydration water under 
surfaces~\cite{BeF96,Be93}.

However, diffraction studies give only indirect information about the 
existence of crystalline or amorphous states in water, because 
the Bragg peaks of ice are quite hard to distinguish from 
liquid states. Moreover, the presence of layers is 
also only obtained from indirect evidences. In order 
to circumvent the difficulties of obtaining the 
structure of water inside the confined system from 
experiments, a number of simulations have been 
performed~\cite{Gi09a,Lo09,Ku05}. They 
employ atomistic models 
such as SPC/E~\cite{Gi06,Gi09a,Lo09} and TIP5P~\cite{Ku05} and 
coarse-grained models~\cite{santos_franzese_2011,St12} for 
water.

Simulations indicate that for both hydrophobic~\cite{Gi06,Gi09,Gi09a,Lo09,Ku05} and 
hydrophilic~\cite{Lo09,Gi06,Gi09a} surfaces
two or three layers are formed depending on the distance
between the confining surfaces. In the case of hydrophobic
walls, there is a phase transition between
the two to the three layers regime
and for a certain temperature
and layer separation the central layer stays liquid while the molecules at
the walls crystallizes.  In addition, in the case 
of hydrophobic walls  the temperature of maximum density
and the temperature of maximum and minimum diffusivity
move to lower temperatures when compared with
the bulk results~\cite{Gi09,Ku05}. At very low pressures, cavitation 
appears~\cite{Gi06}. In the case of the hydrophilic walls, in agreement
with the experimentl results, the system
remains liquid for temperatures below the temperatures in the 
bulk case~\cite{Gi09a}.

Thermodynamic anomalies do
not occur only in water, experiments for $Te$~\cite{Th76}, $Ga$, $Bi$,
$S$~\cite{Sa67,Ke83} and $Ge_{15}Te_{85}$~\cite{Ts91}, liquid metals
\cite{Cu81} and graphite \cite{To97} and simulations for
silica~\cite{An00,Ru06b,Sh02}, silicon~\cite{Sa03} and $BeF_2$~\cite{An00}
shown that these system also have thermodynamic anomalies. In 
addition,  silica~\cite{Sh02, Ch06} and silicon~\cite{Mo05}
show diffusion anomalous behavior. In principle
this systems under confinement could also show a 
shift in the anomalous properties and layering
without having hydrogen bonds. 
Atomistic  and coarse-grained models~\cite{Sa11,St12} for water
are an interesting tool for 
understanding water and its properties, however they 
are not appropriated for seeking for universal 
mechanisms that would be common for water and 
the materials cited above in which the hydrogen bonds are 
not present but still they present the anomalous behavior 
of water.

 Acknowledging that  core softened (CS) potentials may engender 
density and diffusion 
anomalous behavior, a number of CS  potentials were proposed to model 
the anisotropic systems described above.  They possess a repulsive 
core that exhibits a region of softening where the slope changes 
dramatically. This region can be a shoulder or a 
ramp~\cite{He70,Sc00,Bu02,Bu03,Sk04,Fr02,Ba04,Ol05,He05a,He05b,
  Ja98,Wi02,Ma04,Ku04,Xu05,Ol06a,Ol06b,Ol07,Ol08a,Ol08b,Ol09,Gr09,Lo07,Gi07,Fo08,Gr09}.
Despite their simplicity, such models had successfully reproduced 
the thermodynamic, dynamic, and structural anomalous behavior
present in bulk liquid water. They also predict the existence of a second critical 
point hypothesized by Poole and collaborators~\cite{Po92}. This suggests
that some of the unusual properties observed in water can be quite universal and 
possibly present in other systems.

In this work we study the 
effect of the confinement in particles
interacting through a CS potential. 
Our core-softened model introduced
to study bulk system  does not
have any directionality and therefore it is not water. However, 
it does exhibit  the density, the diffusion and
the response functions anomalies observed in water. 
This  suggests that some of the anomalous properties that 
are attributed  directionality of
 water can be found in spherical symmetry systems. 
 We explore that also some of the properties
of water under confinement such as 
the presence of layering  and the shift to lower 
temperatures of maximum density and of maximum and minimum of the
diffusion coefficient can also be obtained with CS potentials.

The paper is organized as follows: in Sec. II we introduce the model; 
in Sec. III the methods and simulation details are described; the 
results are given in Sec. IV; and finally, the conclusions in Sec. V.

\section{\label{sec:model} The Model}

\begin{figure}[!htb]
  \begin{centering}
\includegraphics[clip=true,width=12cm]{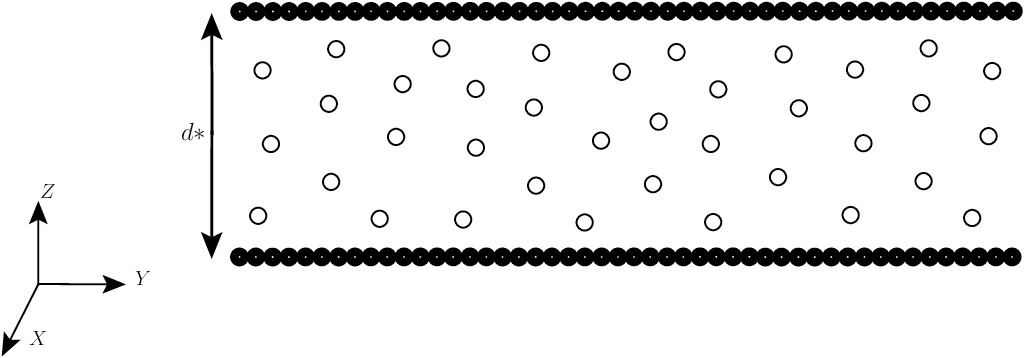}\par
  \end{centering}
  \caption{Model system of particles confined between plates.}
\label{model_figure}
\end{figure}

We study a system of $N$  particles with diameter $\sigma_p$ confined 
between two fixed plates. The surfaces  are formed by particles with 
diameter $\sigma_w$ which are organized in a square lattice of 
area $L^2$.  The center-to-center plates distance is $d^* = d/\sigma_p$. A 
schematic depiction of the system is shown in Fig.~\ref{model_figure}. 

The particles confined between
the two plates interact through an isotropic effective potential given by 
\begin{equation}\label{eq_potential}
 \centering
  \frac{U(r)}{\epsilon} = 4\left[ \left( \frac{\sigma_p}{r} \right)^{12} - 
\left( \frac{\sigma_p}{r} \right)^{6} \right] + 
a\exp\left[-\frac{1}{c^2}\left(\frac{r-r_0}{\sigma_p}\right)^2\right] \;\;.
\end{equation}

The first term is a standard Lennard-Jones (LJ) $12-6$ potential 
with $\epsilon$ depth plus a Gaussian centered on radius 
$r = r_0$ and width $c$. We used parameters $a = 5$, $r_0/\sigma_p = 0.7$ 
and $c = 1$. The pressure versus temperature
phase diagram of this system in the bulk  was studied by 
Oliveira et~al.~\cite{Ol06a, Ol06b}. They found that a
system or particles interacting through this potential
exhibits a region in the pressure-temperature
phase diagram where the density and diffusion coefficient are
anomalous.

\begin{figure}[!htb]
  \begin{centering}
\includegraphics[clip=true,width=9cm]{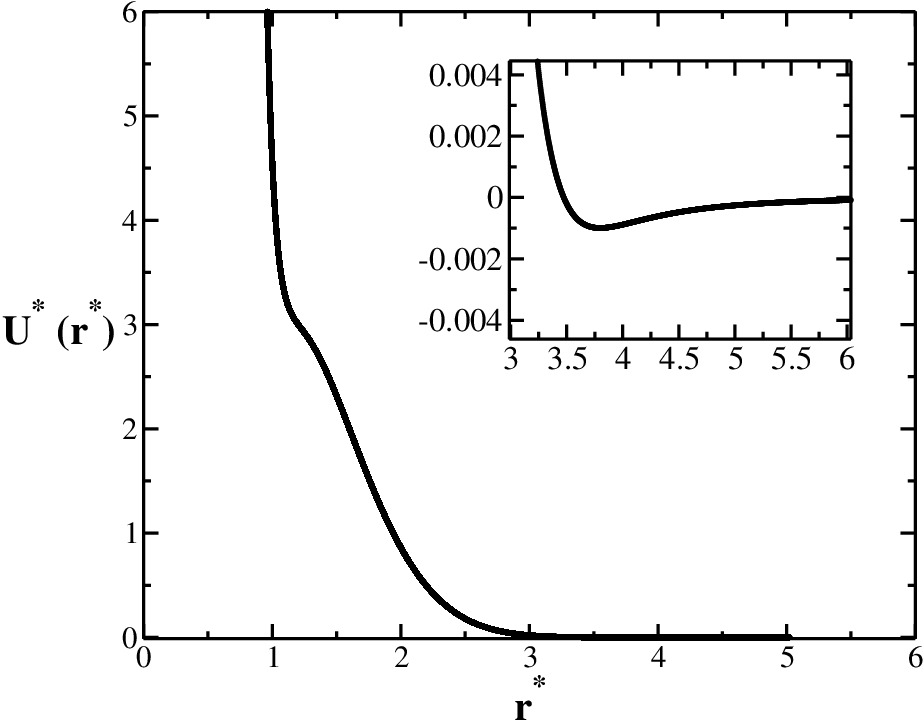}\par
    \par
  \end{centering}
  \caption{Isotropic effective potential Eq.~(\ref{eq_potential}) of 
interaction between the waterlike particles. The potential and the 
distances are in dimensionless units, $U^* = U/\epsilon$ and 
$r^* = r/\sigma_p$ and the parameters are $a = 5$, $r_0/\sigma_p = 0.7$ 
and $c = 1$. The inset shows a zoom in the very small atractive part 
of the potential.}
  \label{potential_alan}
\end{figure}

This potential has two length scales with a repulsive shoulder at  
$r/\sigma_p \approx 1 $ and a very small attractive well at 
$r/\sigma_p \approx 3.8$ (Fig.~\ref{eq_potential}). Depending of the
 choice of the parameters $a,b, c$ and $\sigma_p$, it  can represent a
 whole family of intermolecular interactions. In this paper we employ 
$a = 5$, $r_0/\sigma_p = 0.7$ and $c = 1$.

The particle-plate interaction is given by Weeks-Chandler-Andersen 
Lennard-Jones potential, namely~\cite{WCA_LJ, frenkel}, 
\begin{equation}\label{eq_potential2}
U = \left\{ \begin{array}{ll}
 U_{{\rm {LJ}}}(r) - U_{{\rm{LJ}}}(r_{ cw})  , \qquad r \le r_{ cw} \\
0   , \qquad r  > r_{ cw} \;,
\end{array} \right.
\end{equation}

\noindent where $U_{LJ}(r)$ is a standard 12-6 LJ potential. The
 cutoff distance is $r_{cw} = 2^{1/6}\sigma_{wp}$, where 
$\sigma_{wp} = (\sigma_p + \sigma_w)/2$ is the Lorentz-Berthelot 
mixing rule \cite{Al87} used when two kinds of particles are 
interacting between them. In our model, $\sigma_p = \sigma_w = \sigma_{wp}$. 

\section{\label{sec:simulation} The Methods 
and Simulation Details}

The system has $507$ particles confined between the plates with 
area $L^2$ and 
distant $d$, resulting in a number density $\rho = N/(dL^2)$. The plates 
are located at $z = 0$ and $z = d$, whereas in $x$ and $y$ directions 
periodic boundary conditions are used. The repulsive interactions with 
the plates underestimates the number density, so we need to calculate 
the effective density using the effective distance $d_e$ perpendicular 
to the plates. The new density will be $\rho = N/(d_eL^2)$, where
 $d_e \approx d -(\sigma_p+\sigma_w)/2$ is an approach for the 
effective distance between the plates~\cite{Ku05}. 

Molecular dynamics simulations at the  NVT-constant ensemble
 and the Nose-Hoover~\cite{nose_hoover_85,nose_hoover_86} thermostat were used in 
order to keep fixed the temperature, with coupling parameter $Q = 2$. The 
interaction potential between particles has a cutoff of $r_c = 3.5$ and 
this potential was shifted in order to have $U = 0$ at $r_c$.

Several densities and temperatures are done for the following distances $d^* = d/\sigma_p$
between the plates: 4.2, 4.8, 5.5, 6.0 and 6.3. The initial configuration 
of the systems were set on solid structure and the equilibrium states 
reached after $2 \times 10^6$ steps, followed by $4 \times 10^6$ 
simulation run. The time step was $0.002$ in reduced units and the 
average of the physical quantities were get with $50$ descorrelated 
samples. The thermodynamic stability of the system was checked by 
analyzing the dependence of parallel and perpendicular pressure on 
density namely and by the behavior of the energy after the equilibrium states.

The thermodynamics averages in parallel and perpendicular directions
 to the plates are done employing different
procedures~\cite{kumar_han_stanley_2009}. Parallel 
pressure, $P_{\parallel}$, is computed  using the virial 
expression for the $x$ and $y$ directions~\cite{Ku05, 
Gi09, meyer_stanley_1999, kumar_stanley_2007}, while the 
perpendicular pressure, $P_{\perp}$, is  calculated using two  
distinct methods. For systems with a strong confinement, such as 
$d^* = 4.2$ and $4.8$, the total force perpendicular 
to the plates is used~\cite{meyer_stanley_1999, koga_zeng_tanaka_1998}, 
\begin{eqnarray}
\label{pressao_perpendicular_fa}
P_{\perp} = \frac{F_{plates}}{A} = 
\frac{\left | \sum _{i=1}^{N} {\bf {F}}_{i, plates}\right |}{L^2}\;.
\end{eqnarray} 
For the others systems with larger distances, 
such as $d^* = 5.5$, $6.0$ and $6.3$, the pressure 
$P_{\perp}$ is computed 
through the virial expression in $z$ 
direction~\cite{zangi_rice_2000}.

The dynamic of the systems was studied by lateral diffusion
coefficient, $D_{\parallel}$, related with the mean square 
displacement (MSD) from Einstein relation,
\begin{eqnarray}
\label{difusao_lateral}
D_{\parallel} = \lim_{\tau\to\infty} 
\frac{\langle\Delta r_{\parallel}(\tau)^2\rangle}{4 \tau},
\end{eqnarray}
\noindent where $r_{\parallel} = (x^2+y^2)^{1/2}$ is the distance 
between the particles parallel to the plates. 

We also studied the structure of the systems by lateral radial 
distribution function, $g_{\parallel}(r_{\parallel})$, and 
translational order parameter, $t$. We calculate the 
$g_{\parallel}(r_{\parallel})$ in specific regions between the 
plates, and the same for parameter $t$. An usual definition for 
$g_{\parallel}(r_{\parallel})$ is 
\begin{eqnarray}
\label{gr_lateral}
g_{\parallel}(r_{\parallel}) \equiv \frac{1}{\rho ^2V}
\sum_{i\neq j} \delta (r-r_{ij})\left [ \theta\left( \left|z_i-z_j\right| 
\right) - \theta\left(\left|z_i-z_j\right|-\delta z\right) \right].
\end{eqnarray}
The $\theta(x)$ is the Heaviside function and it restricts the sum of 
particle pairs in the same slab of thickness $\delta z = 1$. We need to 
compute the number of particles for each region and the normalization 
volume will be cylindrical. The $g_{\parallel}(r_{\parallel})$ is 
proportional to the probability of finding a particle at a distance 
$r_{\parallel}$ from a referent particle.

The translational order parameter is defined as \cite{Sh02,Er03,Er01}
\begin{eqnarray}
\label{order_parameter}
t \equiv \int^{\xi _c}_0  \mid g_{\parallel}(\xi)-1  \mid d\xi
\end{eqnarray}
\noindent where $\xi = r_{\parallel}\rho_s^{1/2}$ is the interparticle 
distance in the direction parallel to the plates scaled by 
$\rho_s^{1/2} = (N_{layer}/L^2)^{1/2}$. $N_{layer}$ is the average of 
particles for each slab supposing that this number not change 
significantly (well-defined layers)~\cite{zangi_rice_2000}. We
 use $\xi _c = \rho_s ^{1/2}L/2$ as cutoff distance.

When the system is an ideal gas, with $g_{\parallel}(r_{\parallel}) = 1$, we 
obtain $t = 0$, because the system is not structured. But, as the 
system becomes more structured, like a crystal phase, the 
$g_{\parallel}(r_{\parallel}) \neq 1$, so parameter $t$ assumes large values.

All physical quantities are shown in reduced units~\cite{Al87} as
\begin{eqnarray}
d^* &=& \frac{d}{\sigma_p} \nonumber \\
\tau^* &=& \frac{(\varepsilon/m)^{1/2}}{\sigma _p} \tau \nonumber \\
T^* &=& \frac{k_B}{\varepsilon}T \nonumber \\
P_{\parallel, \perp}^* &=& \frac{ \sigma _p^3 }{\varepsilon}P_{\parallel, \perp} \nonumber \\
\rho^* &=& \sigma _p ^3\rho \nonumber \\
D_{\parallel}^* &=& \frac{(m/\varepsilon)^{1/2}}{\sigma _p}D_{\parallel} \;\; .
\end{eqnarray}

\section{\label{sec:results} Results}

\subsection*{\label{Three-Layers} Systems 
With Three Layers}

\begin{figure}[!htb]
 \centering
 \begin{tabular}{ccc}
 \includegraphics[clip=true,width=5cm]{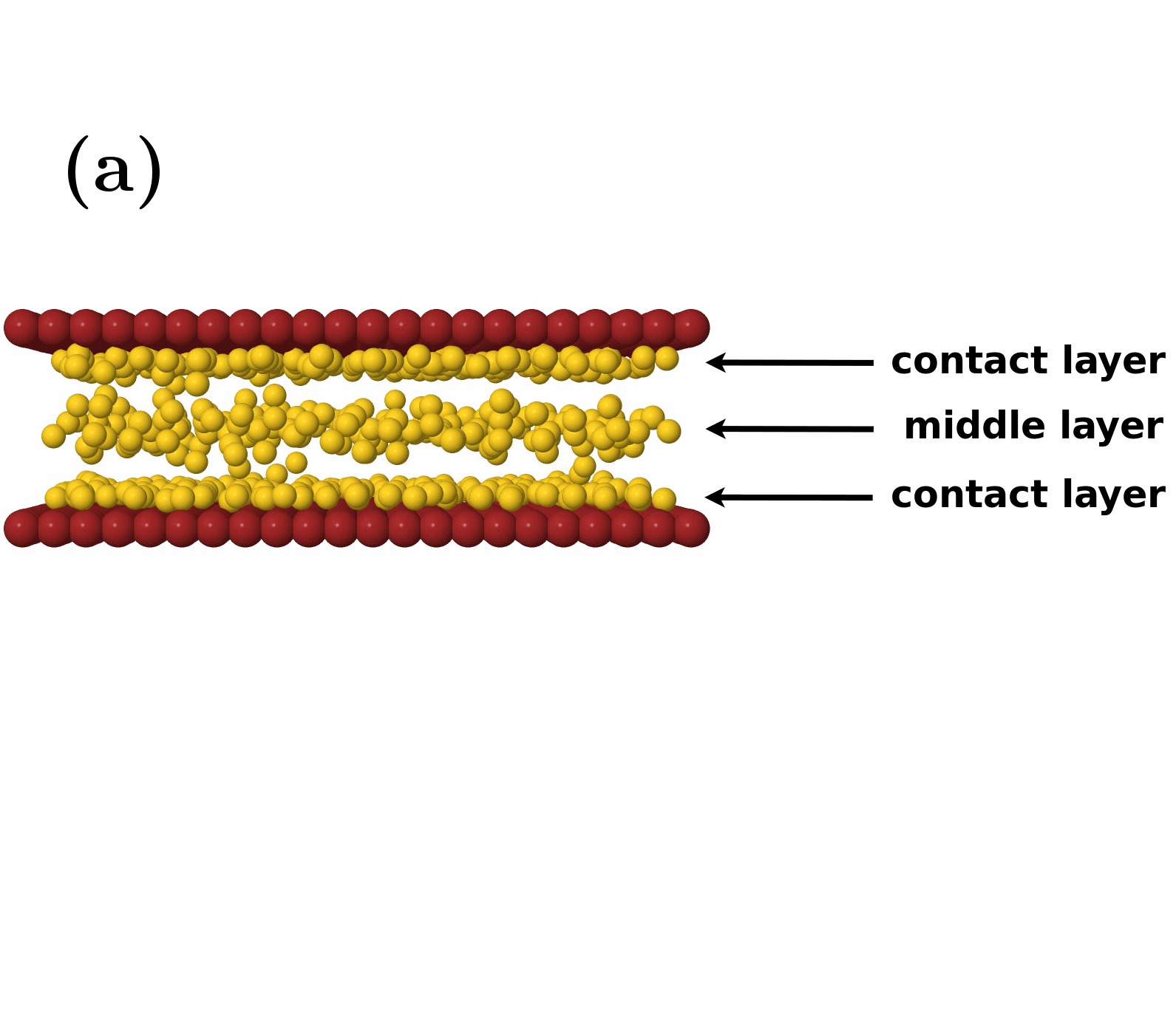} &
 \includegraphics[clip=true,width=5cm]{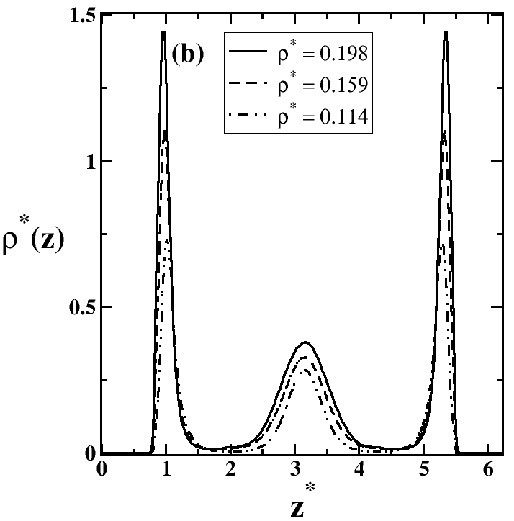} &
 \includegraphics[clip=true,width=5cm]{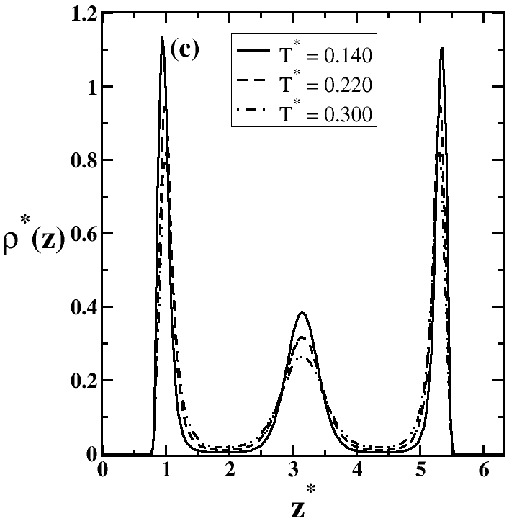} 
\tabularnewline
 \end{tabular}\par
 \caption{In (a), we have a snapshot of the system with
 $T^* = 0.220$ and  $\rho^* = 0.141$ after the equilibrium 
state. Furthermore, the
 transversal density profile is shown for (b) $T^* = 0.220$ and different 
densities, and for 
(c) $\rho^* = 0.141$ and different temperatures. We can see the formation 
of two contact layers, 
near to the plates, and one middle layer. This system corresponds to 
$d^* = 6.3$, whereas the cases 
like $d^* = 5.5$ and $6.0$ present the same behavior in relation to 
layering density.}\label{layering_density}
\end{figure}

The first set of systems that we study corresponds to plates separated by the distances
$d^* = 5.5$, $6.0$ and $6.3$. In all these cases, the particles 
are structured in three layers in $z$ direction divided in two contact layers, near 
to the plates, and one middle layer, located in the center of the plates. The 
formation of layering structures in confined water was also observed 
in atomistic models~\cite{Gi09,Ku05}. The layering 
density can be seen in Fig.~\ref{layering_density} that 
illustrates the $d^* = 6.3$ case: (a) the  snapshot of the system  with
 $T^* = 0.220$ and  $\rho^* = 0.141$, (b) 
the transversal density profile for $T^* = 0.220$ and various densities 
and c) the 
transversal density profile for  
$\rho^* = 0.141$ and different temperatures. The layers become more defined 
at low temperatures and high densities. Now we need to identify if
the different layers are in the solid or in the liquid state. In order
to answer to this question the structure is analyzed.

\begin{figure}[!htb]
 \centering
 \begin{tabular}{cc}
 \includegraphics[clip=true,width=5.5cm]{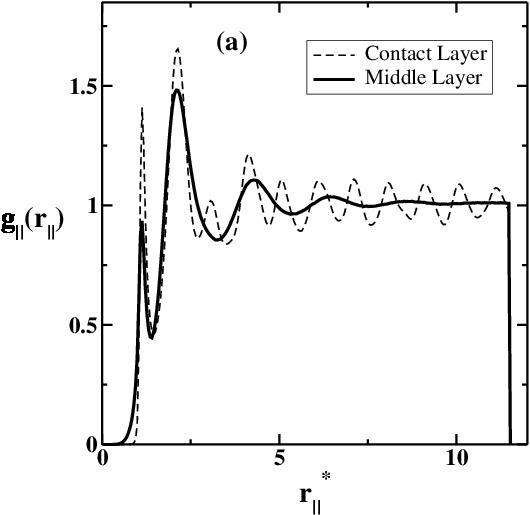} &
 \includegraphics[clip=true,width=5.5cm]{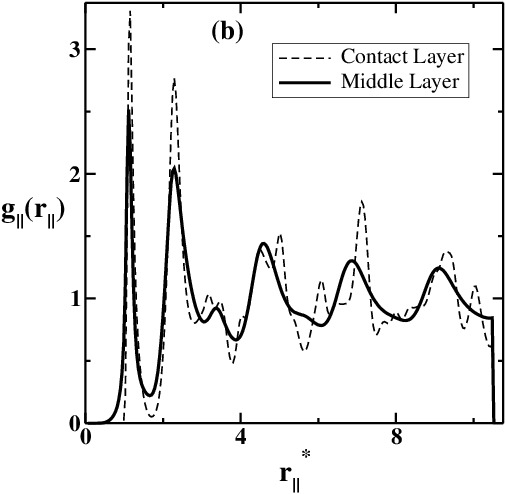} 
\end{tabular}\par
 \caption{Radial distribution function for $\rho^* = 0.181$ and $T^* = 0.220$ in (a),
 and $\rho^* = 0.217$  and $T^* = 0.140$ in (b). Bold lines represent the $g_{||}(r_{||})$ for the middle 
 layer and the dashed lines represent the $g_{||}(r_{||})$ for the contact layer.}\label{gr}
\end{figure}

The Fig.~\ref{gr} shows the radial distribution function for $d^* = 6.3$ in 
two cases: (a) $\rho^* = 0.181$ and $T^* = 0.220$
and (b) $\rho^* = 0.217$ and $T^* = 0.140$. For the case (a) the radial 
distribution of the 
central layer and of the contact layer are liquid-like. The contact 
layer shows a distribution compatible
with a very structured liquid. For the case (b), the central layer is 
also liquid-like, however the 
contact layer is solid-like. The liquid-solid transition occurs at 
different temperatures
and densities for the confinement $d^* = 5.5, 6.0$ and  $d^* = 6.3$ that 
do exhibit
three layers analyzed here. This result is in agreement with 
observations for SPC/E water~\cite{Gi09}.

In the case of the bulk~\cite{Ol06b} the potential exhibit an anomalous
behavior  in the 
translational order parameter $t^*$. For normal systems the $t^*$ grows with 
the density, however for our CS potential it has a region where it does 
decreases 
with the increase of the density. Here, we 
test if this anomalous behavior is also observed in the central and 
contact layers.

\begin{figure}[!htb]
 \centering
 \begin{tabular}{cc} 
 \includegraphics[clip=true,width=5.5cm]{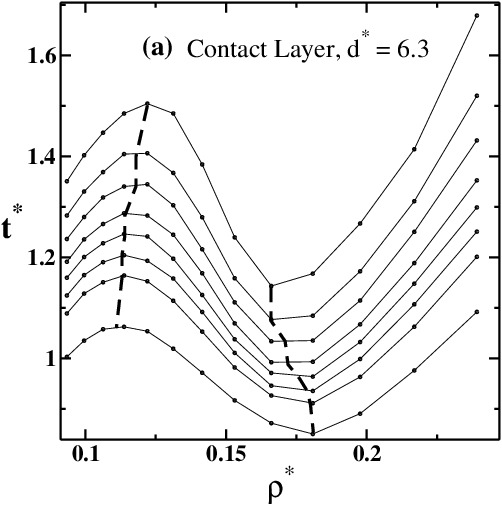} &
 \includegraphics[clip=true,width=5.5cm]{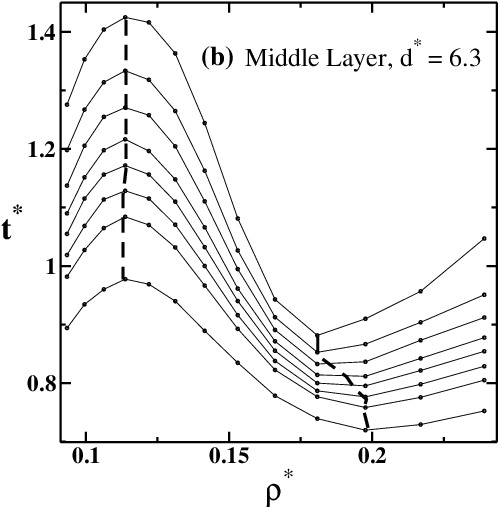} 
\tabularnewline
 \end{tabular}\par
 \caption{ Translational order parameter versus density for $d^* = 6.3$
for (a)  contact layer and the (b) middle layer, separately. The solid lines 
represent the isotherms $T^* = 0.170$, $0.190$, $0.205$, $0.220$, $0.232$, 
$0.245$, $0.260$ and $0.300$ from the top to the bottom. The dashed lines 
connect the extremes.}\label{translational}
\end{figure}

The Fig.~\ref{translational} (a) and (b) show the translational order 
parameter 
defined by Eq.~\ref{order_parameter} as function of density for differents
temperatures $T^*=0.170,0.190,0.205,0.220,0.232,0.245,0.260$ 
and $0.300$, at $d^* = 6.3$, 
for the contact layer and for the middle layer, respectively. The 
dots represent the simulation 
data and the solid lines identify  the isotherms. The
dashed lines, $\rho^*_{t-max} < \rho^* < \rho^*_{t-min}$, identifies
the region in which $t^*$ is anomalous, namely it decreases with the increase of 
the density. The region in
the pressure-temperature phase diagram
in which $t^*$ computed under confinement occurs at lower temperatures
when compared with the bulk values~\cite{Ol06b}.

The translational order parameter for the contact
layer  for very low temperatures
and very high densities is larger than the value for the 
middle layer what suggests that becomes very large the 
indicating the crystallization  the particles at the wall are more structured 
than in the center.
 For the confinement with $d^* = 5.5$ and $6.0$, a similar
behavior for $t^*$ is observed.
The structural differences between the layers present
in our model 
were also observed in water but also in colloidal suspension, for 
several kinds of 
particle-plates interactions \cite{radhakrishnan_gubbins_2000, 
hug_swol_zukoski_1995, 
Ku05, castrillon_aksay_2009, han_kumar_stanley_2009} and for
confined SPC/E water by hydrophobic wall~\cite{Gi09}.

\begin{figure}[!htb]
 \centering
 \begin{tabular}{ccc}
 \includegraphics[clip=true,width=5.5cm]{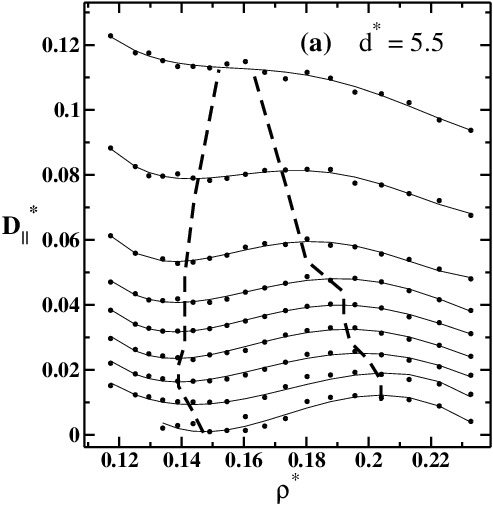} &
 \includegraphics[clip=true,width=5.5cm]{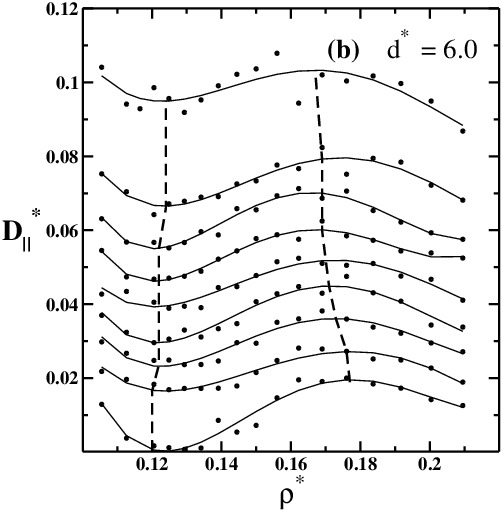} &
 \includegraphics[clip=true,width=5.5cm]{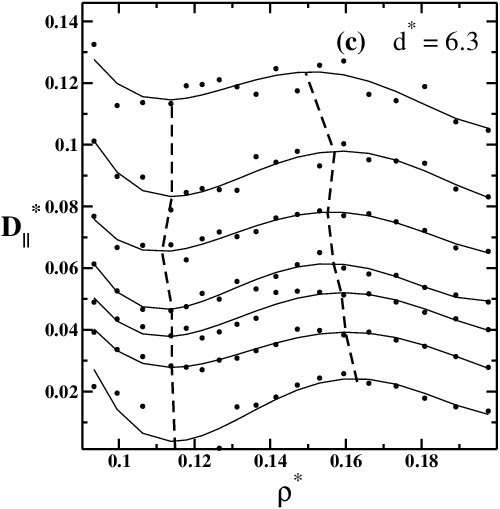} 
\tabularnewline
 \end{tabular}\par
\caption{Lateral diffusion coefficient as function of density for (a) 
$d^* = 5.5$ and isotherms $T^* = 0.140$, $0.160$, $0.175$, $0.190$, 
$0.205$, $0.220$, $0.240$, $0.275$ and $0.320$, (b) $d^* = 6.0$ and 
isotherms $T^* = 0.140$, $0.160$, $0.175$, $0.190$, $0.205$, $0.220$, 
$0.235$, $0.250$ and $0.290$ and (c) $d^*=6.3$ and isotherms $T^* = 0.140$, 
$0.170$, $0.190$, $0.205$, $0.232$, $0.260$ and $0.300$, from the bottom 
to the top. The dots represent the simulation data and the solid line is 
just a polinomial fit to isotherms. The dashed lines connect the extremes 
of diffusion.}\label{diffusion_3}
\end{figure}

Another propertie that exhibits anomalies in the bulk 
is the diffusion coefficient. In normal liquids, the diffusion 
at constant temperature grows with decreasing density, but in 
waterlike liquids 
there is a region 
($\rho^*_{D_{\parallel}-min} < \rho^* < \rho^*_{D_{\parallel}-max}$) 
where the diffusion decreases with decreasing density, so this is an 
anomalous behavior. In 
our system, this anomaly can be observed in the bulk~\cite{Ol06a}. How 
the confinement affects
the region in the pressure-temperature phase diagram where the 
diffusion is anomalous? In order to answer this question 
the lateral diffusion coefficient was computed
as function of density as shown in Fig.~\ref{diffusion_3} 
for (a) $d^* = 5.5$, (b) $d^* = 6.0$ and (c) $d^* = 6.3$. In these cases, 
the diffusion coefficient has a region in which it grows with 
density representing the density anomalous region. The temperature
of maximum and minimum diffusion coefficient are lower in
the confined system than in the bulk case~\cite{Ol06b}.
Our finding are in agreement with 
the observations of the diffusion coefficient
in coarse-grained model confined between smooth 
hydrophobic plates, separated at $0.5$nm~\cite{santos_franzese_2011} 
and atomistic models~\cite{Ku05}.

In the bulk, our potential exhibits a density anomalous region in the 
pressure-temperature
phase diagram. How confining affects the TMD line? In order to answer
to this question the TMD is computed under confinement. 
The density anomaly is given by 
$(\partial \rho /\partial T)_{P_{\parallel}} = 0$, so using the 
following Maxwell relation
\begin{equation}
  \left( \frac{\partial V}{\partial T} \right)_{P_{\parallel}} = -\left(
  \frac{\partial P_{\parallel}}{\partial T} \right)_{V} \left( \frac{\partial
    V}{\partial P_{\parallel}} \right)_{T}, 
\label{tmd}
\end{equation}

\noindent density anomaly can be found through 
$(\partial P_{\parallel}/\partial T)_{\rho} = 0$, what is  equivalent to the 
minimum of the parallel pressure versus temperature.

\begin{figure}[!htb]
 \centering
 \begin{tabular}{ccc}
 \includegraphics[clip=true,width=5.5cm]{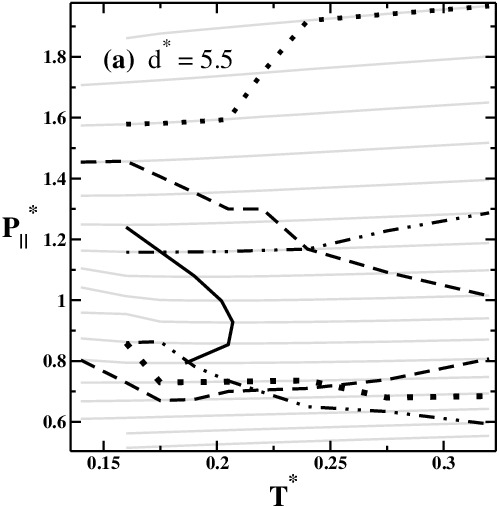} &
 \includegraphics[clip=true,width=5.5cm]{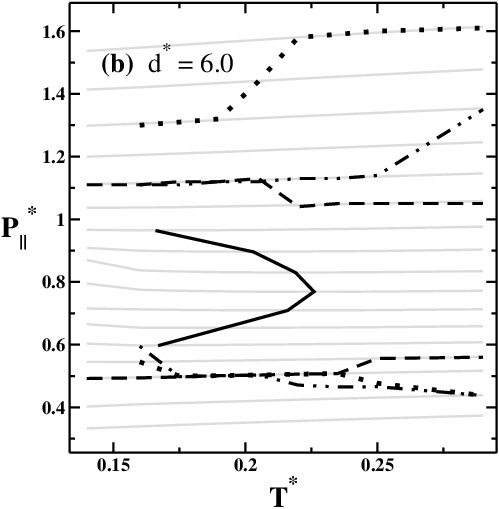} &
 \includegraphics[clip=true,width=5.5cm]{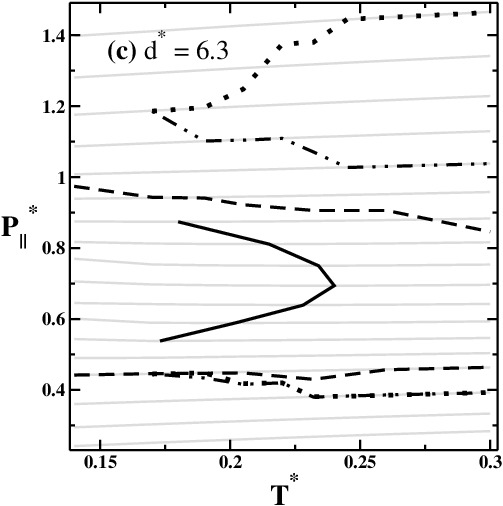} 
\tabularnewline
 \end{tabular}\par
\caption{Phase diagrams $P_{\parallel}^* - T^*$ for (a) $d^* = 5.5$, (b) $d^* = 6.0$ 
and (c) $d^* = 6.3$. The thin solid lines are the isochores (a) 
$0.117 \le \rho^* \le 0.233$, (b) $0.105 \le \rho^* \le 0.209$ and 
(c) $0.099 \le \rho^* \le 0.253$. The solid bold line represents the TMD, the dashed 
line are the lateral 
diffusion extremes, the dashed-dotted line are the translational order parameter 
extremes for a contact layer and the dotted line are the translational order 
parameter extremes for the middle layer.}
\label{phase_diagram_parallel_pressure_T}
\end{figure}

Fig.~\ref{phase_diagram_parallel_pressure_T} illustrates 
the 
parallel pressure versus temperature phase diagram 
for the (a) $d^* = 5.5$, (b) $d^* = 6.0$ and (c) $d^* = 6.3$ cases. The 
thin solid 
lines are the isochores, the solid bold lines represent the TMD, the 
dashed lines are the lateral diffusion 
extremes, the 
dashed-dotted lines are the translational order parameter extremes 
for a contact layer and 
the dotted lines are the translational order parameter extremes for the 
middle layer.

\begin{figure}[!htb]
 \centering
 \begin{tabular}{ccc}
 \includegraphics[clip=true,width=5.5cm]{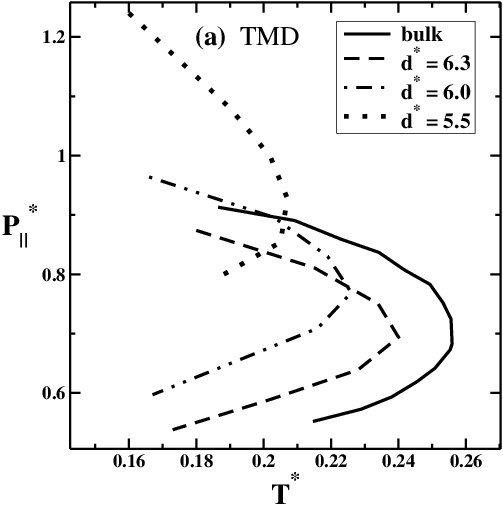} &
 \includegraphics[clip=true,width=5.5cm]{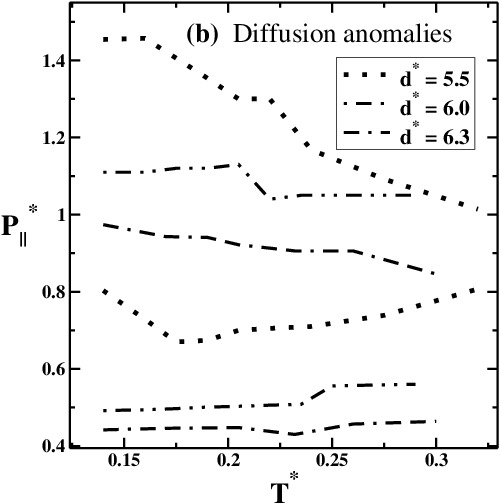} &
 \includegraphics[clip=true,width=5.5cm]{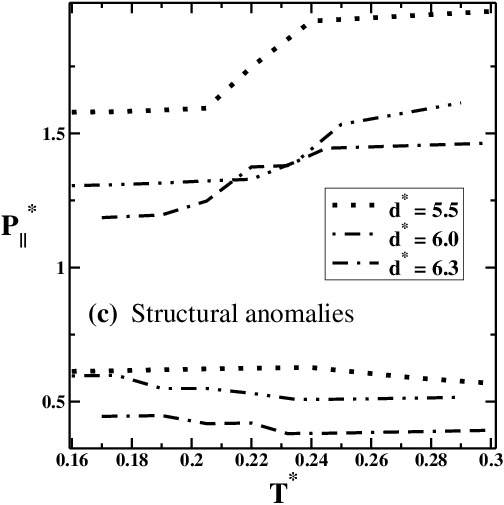} \tabularnewline
 \end{tabular}\par
\caption{Phase diagram $P_{\parallel}^* - T^*$ comparing in (a) the 
shifting of the TMD lines for the confined systems in relation to the
 bulk. A comparison between the anomalies of the confined systems is 
given in (b) for diffusion and in (c) for translational order parameter 
of the middle layer. For (b) and (c), the solid bold lines represent the 
system with $d^* = 5.5$, the dashed-dotted lines represent $d^* = 6.0$ 
and the thin dotted lines represent $d^* = 6.3$. }
\label{phase_diagram_parallel_comparison}
\end{figure}

A comparison between the TMD of the confined
and the bulk systems is given by Fig.~\ref{phase_diagram_parallel_comparison} (a). The 
densities and pressures ranges of the TMD's locations are also shown in 
Table~\ref{table_tmds_3_layers}. The TMD lines of confined systems are 
shifted to lower temperatures and higher densities when compared
with the bulk. 

 For $d^* = 5.5$ and $6.0$, the TMD lines are shifted to higher pressures, 
whereas for $d^* = 6.3$ this shifting occurs to slightly lower pressures when compared
with  the bulk TMD. The non monotonic shift in pressure when
compared with the bulk results can
be attibuted to the fact that we employ the lateral pressure
for the confined system while we use the total pressure for the bulk 
system.

Kumar et~al.~\cite{Ku05} found that the  TMD line for 
confined systems is shifted 
to lower temperatures but in the same range of pressures when
compared with the bulk system . For TIP4P 
water model in contact with six hydrophobic spheres, Gallo and 
Rovere~\cite{gallo_2007} found that  TMD line in confined
systems is shifted to lower temperatures 
and higher pressures. Furthermore, they observed that the spinodal curve 
follow the shifting of the TMD line. Similar result
is observed in our systems. Xu and Molinero~\cite{xu_molinero_2011}, using 
a coarse-grained model for water 
(mW, {\it{Monatomic Water Model}})~\cite{molinero_moor_2009}, confined 
in nanopores of diameter $1.5$nm, also found a TMD shifted to lower 
temperatures and higher pressures. So, the density anomalies observed 
in our systems have a good agreement with other 
atomistic and coarse-grained simulations.

 \begin{table} [!htb]
  \caption{Densities and pressures ranges of TMD's location of 
confined and bulk systems.} 
\vspace{0.5cm}
  \begin{tabular}{c|c|c}
  \hline\hline

\ \ $d^*$       \ \ & \ \    density range           \ \ & \ \   pressure range      \ \ \tabularnewline \hline
\ \ $5.5$       \ & \ \    $0.149 < \rho^* < 0.188$    \ & \ \   $0.800 < P_{\parallel}^* < 1.237$   \ \ \tabularnewline
\ \ $6.0$       \ & \ \    $0.129 < \rho^* < 0.162$    \ & \ \   $0.599 < P_{\parallel}^* < 0.965$   \ \ \tabularnewline
\ \ $6.3$       \ & \ \    $0.122 < \rho^* < 0.153$    \ & \ \   $0.541 < P_{\parallel}^* < 0.875$   \ \ \tabularnewline \hline
\ \ bulk        \ & \ \    $0.110 < \rho^* < 0.140$    \ & \ \   $0.552 < P^* < 0.913$   \ \ \tabularnewline

\hline\hline

  \end{tabular}\label{table_tmds_3_layers}
 \end{table}

\begin{figure}[!htb]
 \centering
 \begin{tabular}{cc}
 \includegraphics[clip=true,width=5cm]{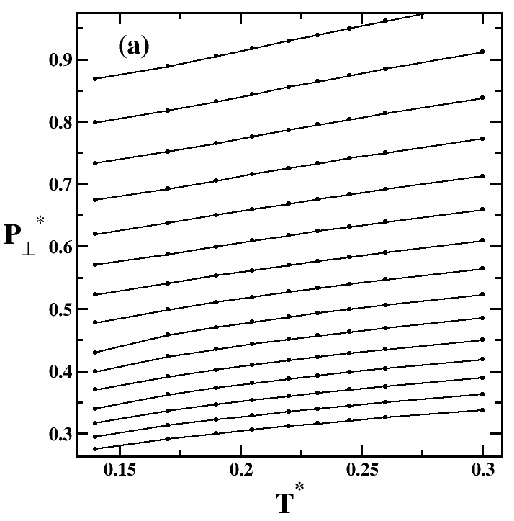} &
 \includegraphics[clip=true,width=5cm]{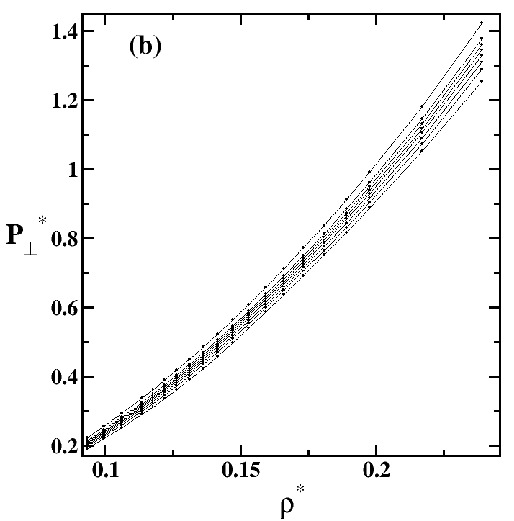} 
\tabularnewline
 \end{tabular}\par
\caption{Phase diagrams (a) $P_{\perp}^* - T^*$ and (b) 
$P_{\perp}^* - \rho^*$ at $d^* = 6.3$. The dots are the simulation data 
and the solid line just connect the isochores in (a) and the isotherms 
in (b). The ranges of densities and temperatures are the same done in 
$P_{\parallel}^* - T^*$ phase diagram.}
\label{phase_diagram_perpendicular}
\end{figure}

The perpendicular pressure is shown in 
Fig.~\ref{phase_diagram_perpendicular} as funtion of (a) 
temperature and (b) density, at $d^* = 6.3$. A monotonic increasing 
behavior is observed in both cases. The ranges of densities and 
temperatures are the same done in $P_{\parallel}^* - T^*$ phase 
diagram. The other systems ($d^* = 5.5$ and $6.0$) have a similar 
behavior and they are not shown here for simplicity. The same behavior 
for perpendicular pressure also was observed by Kumar 
et~al.~\cite{Ku05}.

\subsection*{\label{Two-layers} System 
With Two Layers}

\begin{figure}[!htb]
 \centering
 \begin{tabular}{ccc}
 \includegraphics[clip=true,width=5cm]{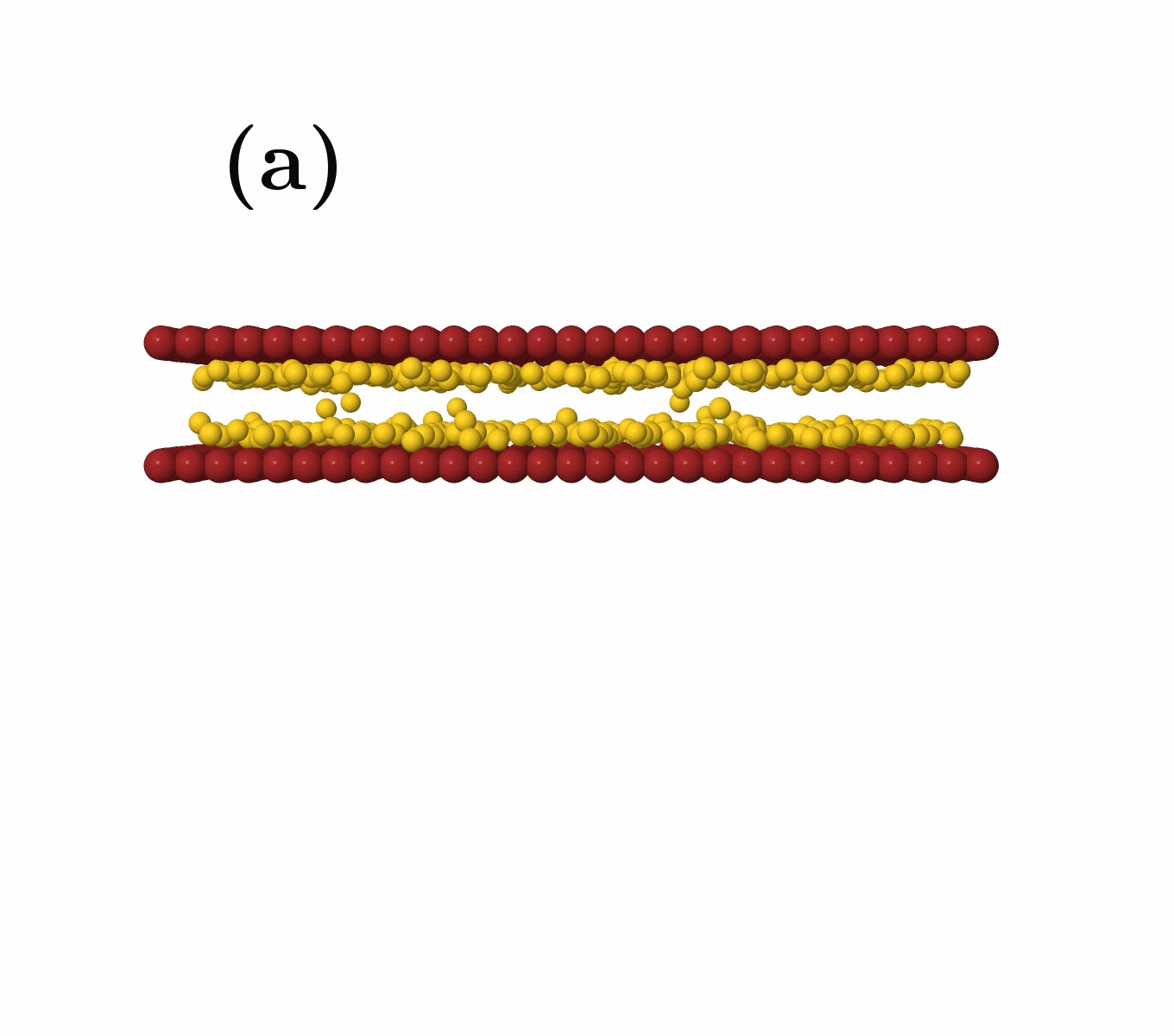} &
 \includegraphics[clip=true,width=5cm]{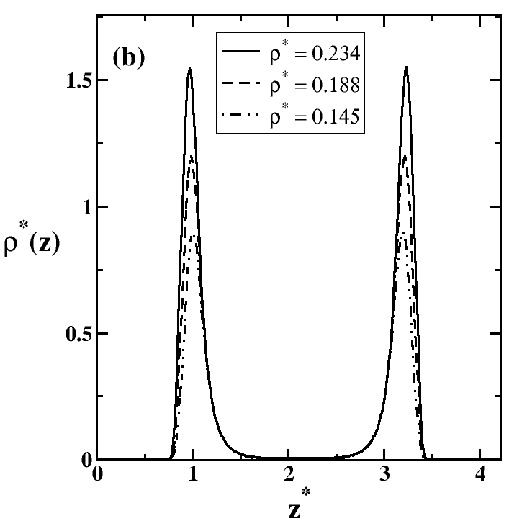} &
 \includegraphics[clip=true,width=5cm]{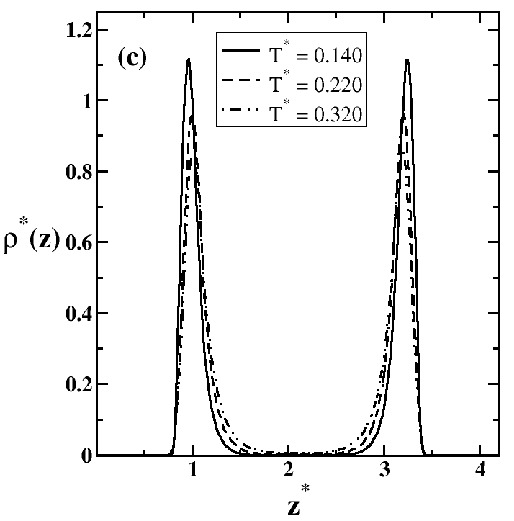} 
\tabularnewline
 \end{tabular}\par
\caption{(a) Snapshot of the system after the equilibrium state. In 
(b) we can see the tranversal density profile for $T^* = 0.220$ and 
several densities, and in (c) for $\rho^* = 0.155$ and several temperatures.}
\label{perfil_4_2}
\end{figure}

The confinement by very narrow distances induces 
the transition from three to two layers.
In this subsection, we study a system with plates separated by $d^* = 4.2$. A 
snapshot   in
Fig.~\ref{perfil_4_2} (a) shows the  two contact layers, 
without middle layer. Figs.~\ref{perfil_4_2} (b) 
illustrate the behavior of the
transversal density profile for fixed 
temperature, $T^* = 0.220$, but for different
total densities. Figs.~\ref{perfil_4_2} (b) also shows the
density profile but for fixed total
density, $\rho^* = 0.155$, and several temperatures.
The structuration in just two contact layers is due the 
strong effect of confinement. Structure of bilayer is observed for 
hydrophobic confinement in TIP5P \cite{zangi_mark_monolayer_2003, 
zangi_mark_bilayer_2003, han_choi_stanley_2010}, 
TIP4P~\cite{koga_1997} and 
mW~\cite{kastelowitz_molinero_2010} models of water.  

\begin{figure}[!htb]
 \centering
 \begin{tabular}{ccc}
 \includegraphics[clip=true,width=5cm]{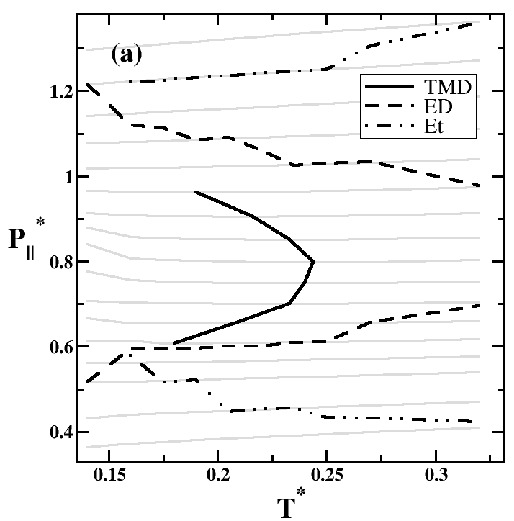} &
 \includegraphics[clip=true,width=5cm]{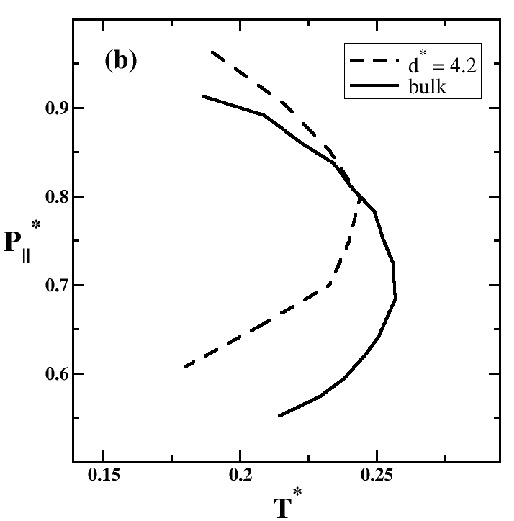} &
 \includegraphics[clip=true,width=5cm]{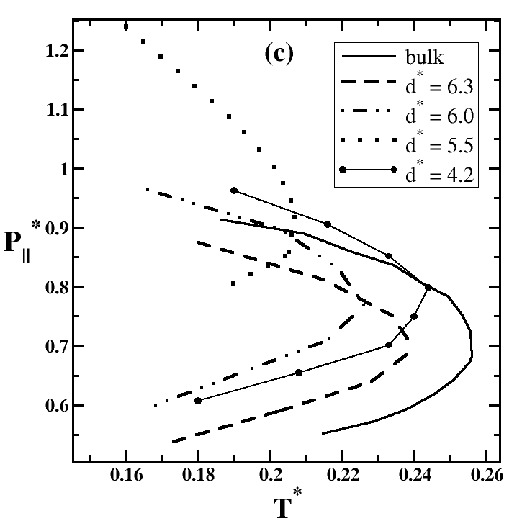} \tabularnewline
 \end{tabular}\par
\caption{Phase diagrams $P_{\parallel}^* - T^*$ showing the density, dynamic 
and structural anomalies in (a). In (b) we have a comparison between the 
TMD lines of $d^* = 4.2$ and the bulk system, and in (c) the same 
comparison is done with TMD lines for systems that present formation of
 three layers.}
\label{diagram_4_2}
\end{figure}

Fig.~\ref{diagram_4_2} (a) shows the parallel pressure 
versus temperature phase diagram. The isochores are represented by thin 
lines, and the TMD by the solid bold line. It
 also shows the  diffusion extrema (dashed line) and the translational 
order parameter extrema (dashed-dotted line). The comparison between the 
TMD line for the $d^* = 4.2$ case and the TMD for the  bulk system is 
illustrated in 
Fig.~\ref{diagram_4_2} (b). The location of the TMD 
is $0.604 < P_{\parallel}^* < 0.959$ and $0.137 < \rho^* < 0.170$. 
Similarly to what happens in the 
three layer system, the TMD line shifts to lower temperatures
when compared with the bulk. This 
characteristic is evidenced in Fig.~\ref{diagram_4_2} (c).

\begin{figure}[!htb]
 \centering
 \begin{tabular}{ccc}
 \includegraphics[clip=true,width=5cm]{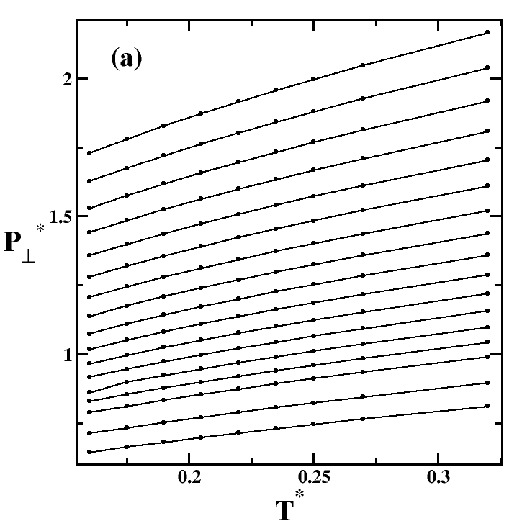} &
 \includegraphics[clip=true,width=5cm]{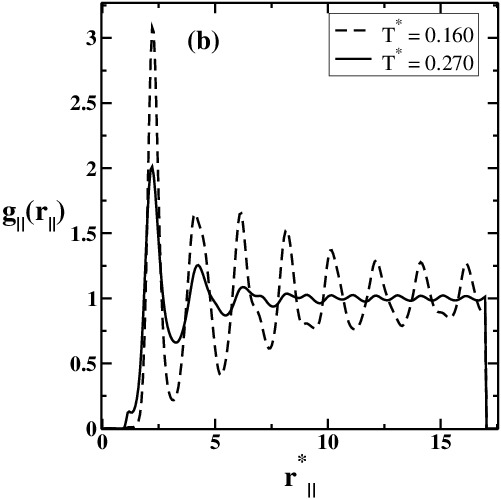} &
 \includegraphics[clip=true,width=5cm]{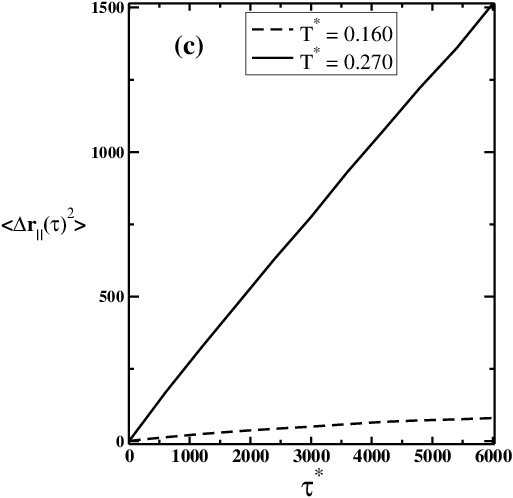} 
\tabularnewline
 \end{tabular}\par
\caption{In (a), the monotonic increasing behavior is observed for 
perpendicular pressure with the temperature. In (b), we have a 
$g_{\parallel}(r_{\parallel})$ for a amorphous solidlike state, with 
$T^* = 0.160$, and a liquidlike state, with $T^* = 0.270$, both at 
$\rho^* = 0.137$. The MSD is observed for these cases in (c).}\label{r2_gr_4_2}
\end{figure}

 Fig.~\ref{r2_gr_4_2} (a) illustrates the 
perpendicular pressure versus temperature phase diagram
for various isochores that show no TMD line. No anomalous
behavior is observed similarly to what happens in the 
three layers regime.
Fig.~\ref{r2_gr_4_2} also shows the radial distribution
function (in (b))  and  the mean square displacement (in (c)) for the lateral 
direction. For $\rho^* = 0.137$, these figures 
show a amorphous solid-like behavior for $T^* = 0.160$ and a liquid-like 
behavior for $T^* = 0.270$. A solid-to-liquid transition
was also observed for the TIP5P 
model~\cite{zangi_mark_bilayer_2003,han_choi_stanley_2010}, for 
the TIP4P model~\cite{koga_1997} and 
for the mW model~\cite{kastelowitz_molinero_2010}.

\begin{figure}[!htb]
 \centering
 \begin{tabular}{cc}
 \includegraphics[clip=true,width=5cm]{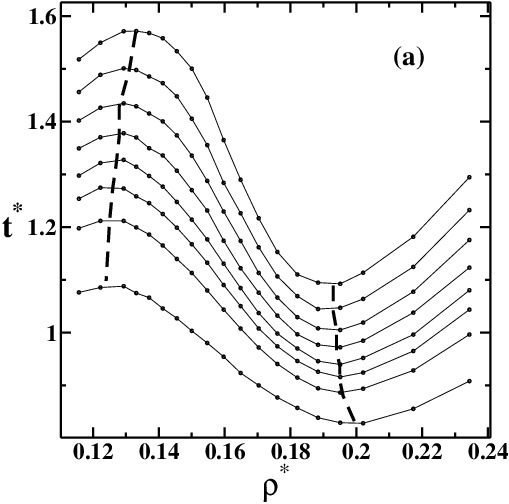} &
 \includegraphics[clip=true,width=5cm]
{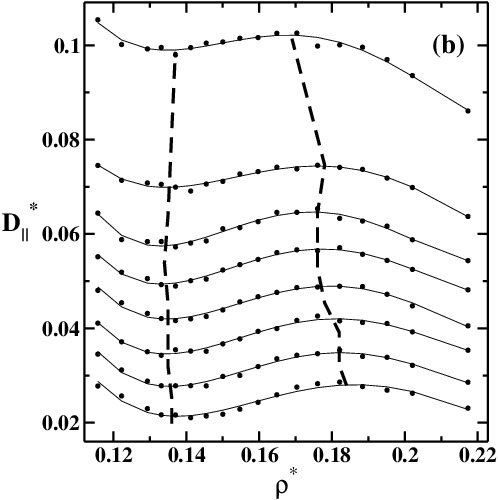} \tabularnewline
 \end{tabular}\par
\caption{In (a), the translational order parameter as function of 
density, and in (b) the lateral diffusion coefficient as function 
of density. The isotherms are connecting the simulation data (points) 
and they are from the top to the bottom $T^* = 0.175$, $0.190$, 
$0.205$, $0.220$, $0.235$, $0.250$, $0.270$ and $0.320$ for $t^*$ (a) 
and from the bottom to the top for $D^*_{\parallel}$ (b). The dashed 
lines connect the extremes observed in both cases.}\label{diff_trans_4_2}
\end{figure}

An anomalous region for translational order parameter and for lateral 
diffusion are  also observed in this extremely confined system
 as can be seen in Figs.~\ref{diff_trans_4_2} (a) and (b), respectively. The 
dashed lines connect the extremes of these anomalies, defining the anomalous 
region that we can see in $P_{\parallel}^* - T^*$ phase diagram in 
Fig.~\ref{diagram_4_2} (a).

\subsection*{\label{phase-diagram} Three-to-Two layers System}

\begin{figure}[!htb]
 \centering
 \begin{tabular}{ccc}
 \includegraphics[clip=true,width=5cm]{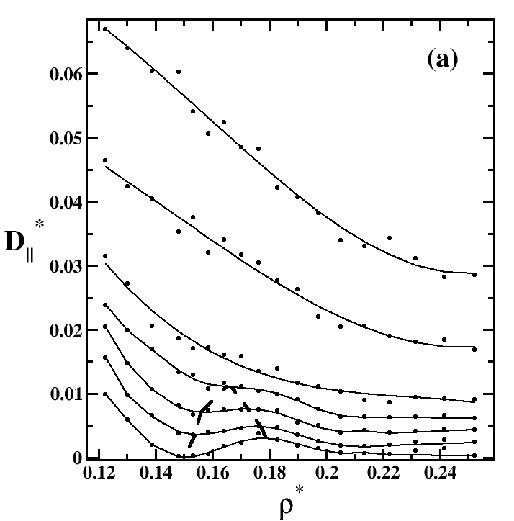} &
 \includegraphics[clip=true,width=5cm]{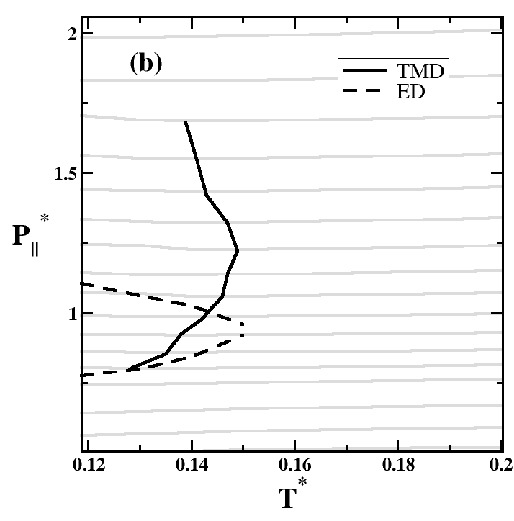} &
 \includegraphics[clip=true,width=5cm]{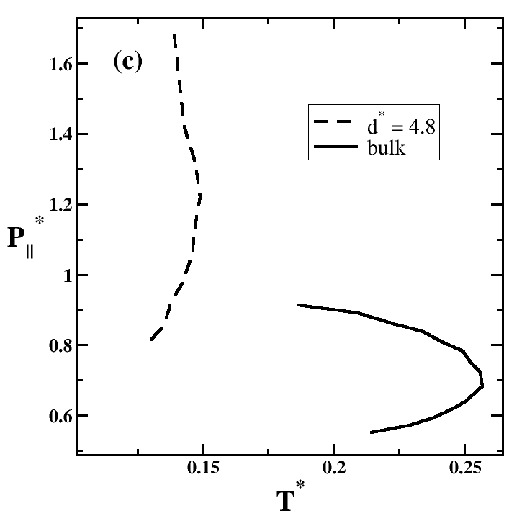} \tabularnewline
 \end{tabular}\par
\caption{In (a), the lateral diffusion coefficient as function of 
density for isotherms from the bottom to the top $T^* = 0.118$, 
$0.130$, $0.140$, $0.150$, $0.160$, $0.190$ and $0.220$. Diffusion 
extremes are connected by dashed lines. In (b), we have the phase 
diagram showing the TMD line and the diffusion extremes. A 
comparison between the TMD line of this system with bulk is given 
in (c).}\label{diagram_diff_4_8}
\end{figure}

The system with $d^* = 4.8$  exhibits
an unusual behavior that resembles properties of both
two layers and the three layers systems. In Fig.~\ref{diagram_diff_4_8} (a) 
the lateral diffusion  as function of density is 
illustrated. The diffusion anomaly is only present at very low 
temperatures,  $T^* \le 0.150$. In addition, as shown in the  
Fig.~\ref{diagram_diff_4_8} (b) the TMD line
is also restricted to low temperatures.  In Fig.~\ref{diagram_diff_4_8} (c) 
the  comparison between the TMD line of the confined system with 
$d^* = 4.8$ and the 
TMD of bulk confirms this shift of the TMD to very
low $T^*$, much lower than the shift observed for confinement with 
$4.8<d^*$ and 
$4.8>d^*$ discussed above.

\begin{figure}[!htb]
 \centering
 \begin{tabular}{ccc}
 \includegraphics[clip=true,width=5cm]{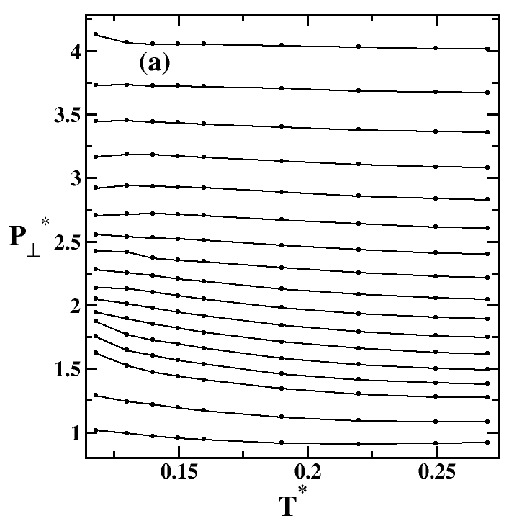} &
 \includegraphics[clip=true,width=5cm]{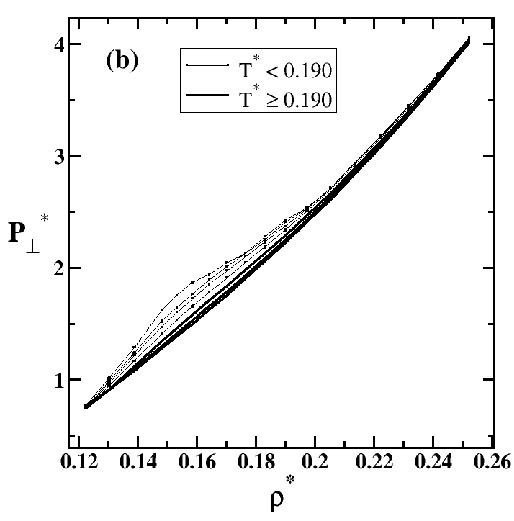} &
 \includegraphics[clip=true,width=5cm]{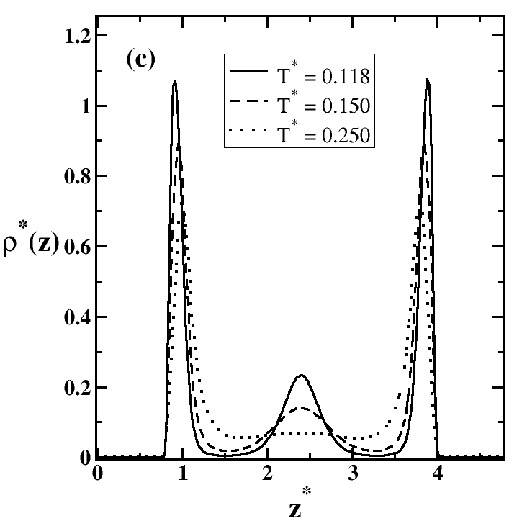} 
\tabularnewline
 \end{tabular}\par
\caption{The perpendicular pressure versus temperature is given in (a) 
and versus density is given in (b). The simulation data (dots) are 
connected by solid lines for better visualization. The transversal 
density profile for $\rho^* = 0.139$ and some temperatures is shown in (c).}
\label{perfil_perp_4_8}
\end{figure}

This $d^*=4.8$ case is not
only peculiar for exhibiting a very low temperature
of maximum density but also for presenting anomalous behavior at the 
perpendicular pressure versus temperature phase diagram.
In Fig.~\ref{perfil_perp_4_8} (a) the isochores 
at the perpendicular pressure versus temperature
phase diagram exhibits  minima not shown for $4.8>d^*>4.8$.
In order to shade some light in the reason for
the unusual behavior of the system for $d^*=4.8$ we explore 
the behavior of the  structure and of the stability of the layers. 
 Fig.~\ref{perfil_perp_4_8} (b) shows the lateral pressure versus
density of fixed temperatures. For $T^* < 0.190$, the slope of 
the curve first increases and then decreases. This change
even thougth not indicating a phase transition is usually
observed before the phase separation would be stablished~\cite{Gi09a}.

The Fig.~\ref{perfil_perp_4_8} (c) shows the tranversal density profile 
for $\rho^* = 0.139$ and for a number of temperatures. At this density, 
three layers are 
present  for low temperatures, $T^* = 0.118$ and $0.150$. At high 
temperatures, 
$T^* = 0.250$, two layers are well defined with particles equally distributed 
between
them without forming a third layer.

\begin{figure}[!htb]
 \centering
 \begin{tabular}{ccc}
 \includegraphics[clip=true,width=5cm]{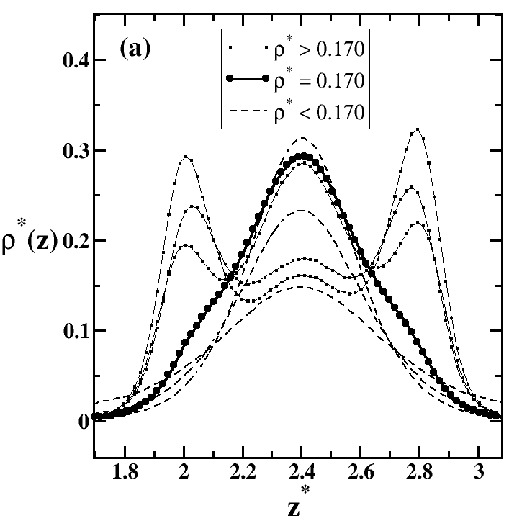} &
 \includegraphics[clip=true,width=5cm]{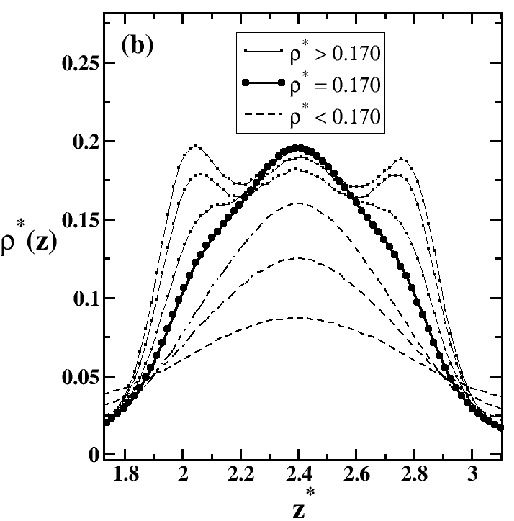} &
 \includegraphics[clip=true,width=5cm]{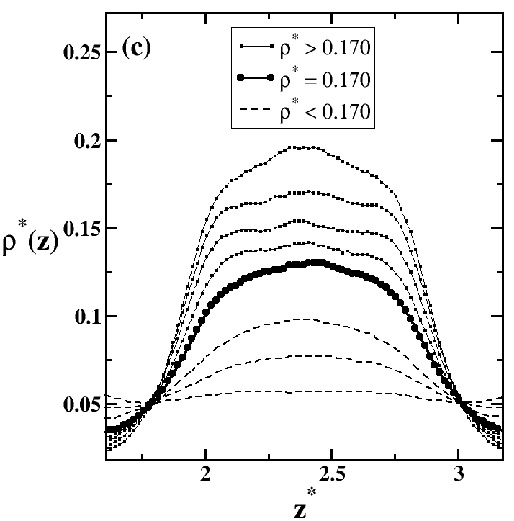} 
\tabularnewline
 \end{tabular}\par
\caption{Tranversal density profile for the middle layer at many 
densities and (a) $T^* = 0.118$, (b) $T^* = 0.160$ and (c) $T^* = 0.220$. The 
dashed lines represent states for $\rho^* < 0.170$, the solid lines 
for $\rho^* > 0.170$ and the bold circles are for $\rho^* = 0.170$.}
\label{transversal_density_profile_4_8}
\end{figure}

In order to understand the two-to-three layers
transition, the change in the 
structure of the central layer with temperature
and density is checked. The structure of the middle layer is shown in
the  Fig.~\ref{transversal_density_profile_4_8} for different
temperatures
and densities.  At low temperatures (cases (a) at $T^*=0.118$ and (b) at $T^*=0.160$) and high densities, 
$\rho^*>0.170$,
the central layer is divided in many sublayers. By decreasing
the density the many layers give rise at $\rho^*=0.170$ 
to a central layer that disappears as the 
density is decreased any further. At high temperatures, $T^*>0.160$ (case (c) shown at $T^*=0.220$), 
as the density decreases the system passes from three-to-two layers
without forming the sublayers. At low temperatures, due to
the presence of the many sublayers the density anomalous 
region appears. The TMD  originates from particles moving from
one scale to the other~\cite{Ol06b}.

Our interpretation 
is in accordance with
the assumption of  Kumar et~al.~\cite{Ku05}, that
the split of the middle 
layer in sublayers is justified by the density anomaly.  Using SPC/E 
model for water and hydrophobic rough 
plates, Giovambattista et~al.~\cite{Gi09} 
found a phase transition between a bilayer ice and a trilayer 
heterogeneous fluid for different distances between the plates. For 
smooth plates separated at $0.8$nm and SPC/E model, Lombardo 
et~al.~\cite{Lo09} also found a phase 
transition between two and three layers. 

Changes in diffusion coefficient as function of separation 
between the plates are reported in systems with transitions 
between 6 and 5, 5 and 4, 4 and 3 layers \cite{gao_1997}, suggesting 
that the dynamic behavior change in systems with structural
 transitions. So, the dynamic behavior of our system at $d^* = 4.8$ is another 
possible explanation for the peculiar behavior on its anomalies.


\section{\label{sec:conclusions} Conclusions}

In this paper, the effects of confining a system of 
particles interacting through core-softened is explored.

The formation of three layers, two close to the walls and 
one central is observed for large values of $d^*$, while
two layers are observed for small values of $d^*$.
In addition the region in the pressure-temperature 
phase diagram where the density anomaly appears moves to
lower temperatures.
These results are similar to the 
results obtained in atomistic and 
coarse-grained models where unlike our
model the directionality of the h-bonds is
explicitly included~\cite{Ku05, 
han_diffusion_perpendicular_2008}.

Our results indicate that layers are formed in
order to minimize the particle-particle interaction
potential and the wall-particle interaction. Therefore,
if the walls are distance $d^*=5.0$ it is possible
to fit three layers in which each will is distant 
$2.0$ from the other and the contact layer is 
distant $1.0$ from the wall. For $d^*=4.0$ it is
only possible to fit two layers distant  $2.0$ from
each other and $1.0$ from the wall. The density, diffusion
and structural anomalous behavior that implies particles
moving from one length scales to the other (moving 
from the length scale at $\approx 3.0$ to the 
length scales at $1.0$) occurs only along the 
parallel plane, therefore the anomalies appear 
as function of $P_{\parallel}$.

The case $d^*=4.8$ is the boundary between the 
two layer and the three layer cases. This case
allow us to observe how the presence of 
the density anomaly is related with moving from
different particle-particle distances.

Our results suggest that 
effective spherical symmetric two length 
scales potentials are an interesting 
tool for understanding the mechanisms that 
arise from confining systems with density,
diffusion and structural anomalies. Due to
their simplicity the results obtained 
can be generalized to other experimental realizations 
besides water.

\section*{ACKNOWLEDGMENTS}

We thank for financial support the Brazilian science agencies, CNPq 
and Capes. This work is partially supported by CNPq, INCT-FCx. We also 
thank to CEFIC - Centro de F\'isica Computacional of Physics Institute 
at UFRGS, for the computer clusters.

\vspace{1cm}
\bibliography{Biblioteca-leandro-new}

\end{document}